\documentclass[twocolumn,trackchanges]{aastex631}

\shorttitle{Igneous rim accretion in slow shock}
\shortauthors{Matsumoto \& Arakawa}

\newcommand{\erf}[1]{{\rm erf}{{#1}}}
\newcommand{\We}{{\rm We}}
\newcommand{\Ca}{{\rm Ca}}

\begin{document}

\title{Igneous Rim Accretion on Chondrules in Low-Velocity Shock Waves}

\correspondingauthor{Yuji Matsumoto}
\email{yuji.matsumoto@nao.ac.jp}

\author[0000-0002-2383-1216]{Yuji Matsumoto}
\affiliation{National Astronomical Observatory of Japan, 2-21-1, Osawa, Mitaka, 181-8588 Tokyo, Japan}
\author[0000-0003-0947-9962]{Sota Arakawa}
\affiliation{Japan Agency for Marine-Earth Science and Technology, 3173-25, Showa-machi, Kanazawa-ku, Yokohama, Kanagawa 236-0001, Japan}

\begin{abstract}
Shock wave heating is a leading candidate for the mechanisms of chondrule formation.
This mechanism forms chondrules when the shock velocity is in a certain range. 
If the shock velocity is lower than this range, dust particles smaller than chondrule precursors melt, while chondrule precursors do not.
We focus on the low-velocity shock waves as the igneous rim accretion events.
Using a semi-analytical treatment of the shock-wave heating model, we found that the accretion of molten dust particles occurs when they are supercooling. 
The accreted igneous rims have two layers, which are the layers of the accreted supercooled droplets and crystallized dust particles.
We suggest that chondrules experience multiple rim-forming shock events.
\end{abstract}

	
\section{Introduction} \label{sect:intro}

Chondrules are a major ingredient of chondritic meteorites (up to 80\% in volume).
They are roughly millimeter-sized spherical igneous grains \citep[e.g.,][]{Scott2007}.
Their spherical and igneous features are reproduced when their precursor grains were molten through flash heating events and cooled in the solar nebula \citep[e.g.,][]{Hewins+2005}. 
Chondrules are often surrounded by rims.
Rims are divided into two major groups: fine-grained rims and igneous rims. 
Fine-grained rims resemble the matrix and are considered as the formation by the accretion of fine-grained dust onto chondrules \citep[e.g.,][]{Ashworth1977, Hanna&Ketcham2018}.
Igneous rims are composed of coarse grains and show evidence of appreciable melting during flash heating events similar to chondrule formation ones \citep[e.g.,][]{Rubin1984,Matsuda+2019}.
These rims provide information about their accretion and heating events, which are additional information about chondrule-forming environments.

The formation process of igneous rims is often interpreted as the melting of preexisting fine-grained rims \citep[e.g.,][]{Rubin1984,Rubin2010}.
This interpretation is consistent with multiple melting events of chondrules \citep[e.g.,][]{Rubin&Wasson1987, Wasson1993, Baecker+2017}.
However, this does not provide the explanation for the larger thicknesses of igneous rims compared to fine-grained rims \citep{Matsumoto+2021}, the isotopic relations between igneous rims and their host chondrules \citep{Jacquet+2015}, or the association between igneous rims and fine-grained rims \citep{Pinto+2022}.
Additional mechanisms would work to form igneous rims, such as the accretion of droplets \citep{Jacquet+2013, Jacquet+2015}.

A plausible event for droplet formation is heating events similar to chondrule formation events.
Several mechanisms have been proposed for the chondrule formation events, such as shock waves \citep[e.g.,][]{Hood&Horanyi1991,Iida+2001}, planetesimal collisions \citep[e.g.,][]{Asphaug+2011, Johnson+2015, Wakita+2017}, and lightning \citep[e.g.,][]{Horanyi+1995, Johansen&Okuzumi2018}.
\cite{Miura&Nakamoto2005} reported that $10~\mu{\rm m}$ sized dust particles are molten and become droplets in the shock wave heating events when the shock speed is lower than that in chondrule formation events.
This size is the minimum one of the dust particles that can survive in shock wave heating events, and dust particles smaller than this size evaporate away completely \citep[see also][]{Jacquet&Thompson_C2014}.
If chondrules accrete these droplets in the postshock region, the droplets form igneous rims.

\begin{figure}[ht!]
    \plotone{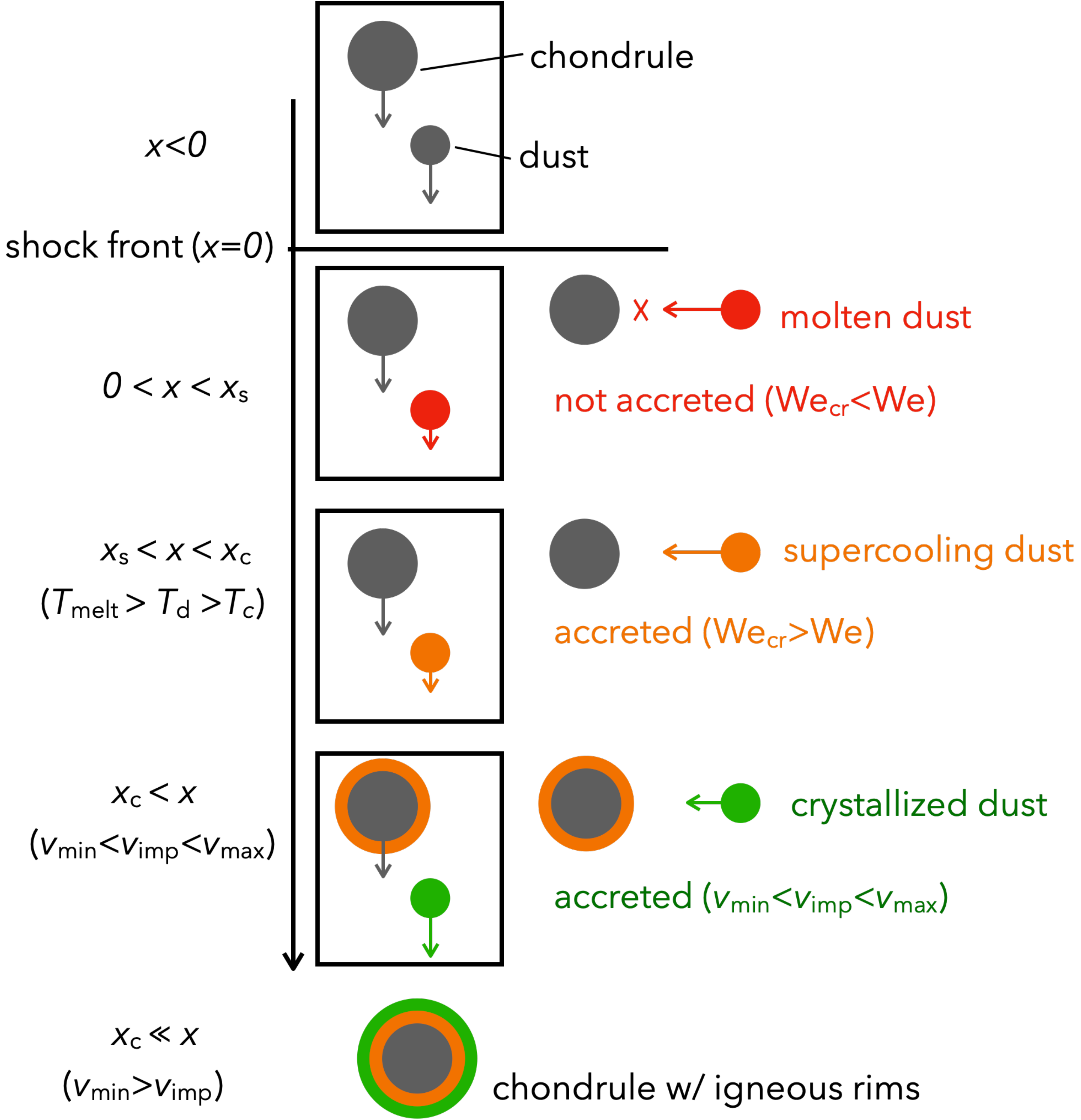}
    \caption{Schematic figure of the igneous rim accretion behind the shock front.
    In the preshock region, $x<0$, both chondrules (the larger circle) and dust particles (the smaller circle) are comoving with gas.
    Behind the shock front, dust particles are heated and molten due to the gas drag. 
    A molten dust particle can be accreted onto a chondrule when the Weber number ($\We$) is smaller than the critical Weber number ($\We_{\rm cr}$).
    Molten dust particles are not accreted onto chondrules due to high impact velocities ($\We>\We_{\rm cr}$) at $0<x<x_{\rm s}$.
    Molten dust particles turn into a supercooled state when the dust temperature, $T_{\rm d}$, becomes lower than the melting temperature, $T_{\rm melt}$, at $x=x_{\rm s}$.
    Supercooling dust particles can be accreted onto chondrules after their viscosities become higher and they satisfy $\We<\We_{\rm cr}$.
    Dust particles are crystallized when $T_{\rm d}$ becomes the crystallization temperature, $T_{\rm c}$, at $x_{\rm c}$. 
    Crystallized dust particles are accreted onto chondrules when the impact velocity satisfies the accretion condition, $v_{\rm min}<v_{\rm imp}<v_{\rm max}$.
    After dust particles and chondrules become comoving with gas, their relative velocity becomes lower than $v_{\rm min}$, and the igneous rim accretion behind the shock front finishes.
    Finally, chondrules obtain two-layered igneous rims, which are formed via the accretion of molten dust particles and crystallized dust particles, respectively.
    In both layers, rim particles experienced melting.
    \label{fig:schematic}}
\end{figure}

In this paper, we have investigated the accretion of igneous rims in the postshock regions.
We consider the dynamical and thermal evolution of chondrules and dust particles behind the shock front and evaluate whether molten droplets are accreted onto chondrules.
Our model is described in Section \ref{sect:model}.
We find the igneous rims are accreted behind the shock front. 
Figure \ref{fig:schematic} is a schematic figure of igneous rim accretion.
We show how rims are accreted behind the shock front in Section \ref{sect:results}.
In Section \ref{sect:conclusion}, we summarize our results and discuss the accretion of the igneous rims.

\section{Model} \label{sect:model}

The dynamical and thermal evolution of chondrules behind the shock front are investigated by the studies of chondrule collisions \citep[e.g.,][]{Nakamoto&Miura2004_LPSC, Ciesla2006, Jacquet&Thompson_C2014, Arakawa&Nakamoto2019} and the accretion of fine-grained rims around chondrules \citep{Arakawa+2022} using one-dimensional normal shock models.
In this paper, we adopt the one-dimensional normal shock model, which is based on \cite{Arakawa&Nakamoto2019} and \cite{Arakawa+2022}, and modified it to roughly reproduce the results in \cite{Miura&Nakamoto2005}.

\subsection{Gas Structure}

We assume a simple gas structure. 
In our model, the gas velocity with respect to the shock front, $v_{\rm g}$, is expressed as a function of the distance from the shock front, $x$:
\begin{eqnarray}
    v_{\rm g} = 
    \left\{
        \begin{array}{l}
            v_0 \hfill(x<0),\\
            v_0 + (v_{\rm post} -v_0) \exp{(-x/L)}\quad \hfill (x\geq0),
        \end{array}
      \right.
      \label{eq:v_g}
\end{eqnarray}
where $v_0$ is the pre-shock gas velocity with respect to the shock front, namely, shock velocity, $v_{\rm post}$ is the post-shock gas velocity with respect to the shock front, and $L$ is the spatial scale of the shock.
The post-shock gas velocity is given by the Rankine--Hugoniot relations as $v_{\rm post}= [(\gamma-1)/(\gamma+1)]v_0$, where $\gamma$ is the ratio of specific heats, $\gamma=1.4$.
We take $L$ as a parameter and its fiducial value is $L=10^3$~km \citep{Miura&Nakamoto2005}. 
The temperature of the gas, $T_{\rm g}$, is 
\begin{eqnarray}
    T_{\rm g} = 
    \left\{
        \begin{array}{l}
            T_0 \hfill(x<0),\\
            T_0 + (T_{\rm post} -T_0) \exp{(-x/L)}\quad \hfill (x\geq0),
        \end{array}
      \right.
\end{eqnarray}
where the preshock gas temperature is $T_0=500$~K and the post-shock gas temperature is $T_{\rm post}=2000$~K.
The postshock region is approximately isobaric \citep[e.g.,][]{Susa+1998,Iida+2001}.
According to the Rankine--Hugoniot relation, the gas number density in the postshock region, $n_{\rm g}$, can be expressed by
\begin{eqnarray}
    n_{\rm g} = n_0 \frac{T_0}{T_{\rm g}} \frac{ 4 s_0^2 - (\gamma-1)}{\gamma+1},
    \label{eq:n_g}
\end{eqnarray}
where $n_0$ is the preshock gas number density, $s_0=v_0/(2k_{\rm B} T_{0}/m_{\rm g} )^{1/2}$, $k_{\rm B}$ is the Boltzmann constant, and $m_{\rm g}=3.34\times 10^{-24}$~g is the gas molecule mass.

\subsection{Dynamical and Temperature Evolution of Chondrules and Dust}

In the postshock region, the velocities of chondrules and dust particles with respect to the shock front are given by 
\begin{eqnarray}
    m \frac{{\rm d} v}{{\rm d} x} = -\frac{ C_{\rm D} }{2} \pi a^2 m_{\rm g} n_{\rm g} \frac{|v-v_{\rm g}|}{v} (v-v_{\rm g}),
\end{eqnarray}
where $v$, $m$, and $a$ are the velocity, mass, and radius of a chondrule or particle.
We consider the radii of chondrules, $a_{\rm ch}$, and initial dust particles,  $a_{\rm d,init}$, as parameters, and their fiducial values are $a_{\rm ch}=10^3~\mu$m and $a_{\rm d,init}=20~\mu$m.
The densities of the chondrules and the dust particles are the same, $\rho_{\rm ch}=\rho_{\rm d}=3.3~\mbox{g~cm}^{-3}$, whose values are taken from the typical values for chondrules \citep[e.g.,][]{Hughes_DW1978, Friedrich+2015}.
This density is close to that of forsterite, from which we take the other material properties in this paper.
The drag coefficient, $C_{\rm D}$, is given by \citep{Arakawa+2022}
\begin{eqnarray}
    C_{\rm D} &=& \frac{2}{3s}\sqrt{\frac{\pi T}{T_{\rm g}} } + \frac{2s^2+1}{\sqrt{\pi} s^3} \exp{(-s^2)} 
    + \frac{4s^4+2s^2-1}{2s^4} \erf{(s)}\nonumber\\
    &\simeq&
    \frac{2\sqrt{\pi}}{3s} + \frac{2s^2+1}{\sqrt{\pi} s^3} \exp{(-s^2)} 
    + \frac{4s^4+2s^2-1}{2s^4} \erf{(s)}.
    \nonumber\\
\end{eqnarray}
where $s = v /(2k_{\rm B} T_{\rm g}/m_{\rm g} )^{1/2}$.
This approximation works well when $T\simeq T_{\rm g}$ or $s>1$.
Either condition is satisfied behind the shock front when dust particles are molten.

The temperatures of chondrules and dust are derived by the equations of energy, 
\begin{eqnarray}
    m c_{\rm heat} \frac{{\rm d} T}{{\rm d} x} =
    \frac{4\pi a^2}{v} (\Gamma - \Lambda_{\rm rad} - \Lambda_{\rm evap} ),
\end{eqnarray}
where $c_{\rm heat}=1.42\times 10^7~\mbox{erg~g$^{-1}$~K$^{-1}$}$ is the specific heat, $\Gamma$ is the heating rate via the energy transfer from gas, $\Lambda_{\rm rad}$ is the radiative cooling, and $\Lambda_{\rm evap}$ is the latent heat cooling by evaporation per unit area, respectively \citep[e.g.,][]{Miura+2002,Miura&Nakamoto2005}.
The heating rate via the energy transfer from gas is given by
\begin{eqnarray}
    \Gamma = m_{\rm g} n_{\rm g} | v-v_{\rm g} | (T_{\rm rec} -T) C_{\rm H},
    \label{eq:Gamma}
\end{eqnarray}
where the recovery temperature, $T_{\rm rec}$, and the heat transfer function, $C_{\rm H}$, are \citep[e.g.,][]{Gombosi+1986}
\begin{eqnarray}
    T_{\rm rec} &=& \frac{T_{\rm g}}{\gamma+1} 
    \left[ 2\gamma+ 2(\gamma-1) s^2 \right. \nonumber\\ &&\left.
    - \frac{\gamma-1}{ 0.5+ s^2 + (s/\sqrt{\pi}) \exp{(-s^2)} {\rm erf}^{-1}(s) } \right],
\end{eqnarray}
and
\begin{eqnarray}
    C_{\rm H} &=& \frac{\gamma+1}{\gamma-1} \frac{ k_{\rm B} }{8m_{\rm g} s^2} 
    \left[ \frac{s}{\sqrt{\pi}}\exp{(-s^2)} + \left( \frac{1}{2}+s^2\right)\erf{(s)} \right],
    \nonumber\\
\end{eqnarray}
respectively.
The radiative cooling rate is given by
\begin{eqnarray}
    \Lambda_{\rm rad} &=& \epsilon_{\rm emit} \sigma_{\rm SB} T^4 - \epsilon_{\rm emit} \sigma_{\rm SB} T_0^4,
    \label{eq:Lambda_rad}
\end{eqnarray} 
where $\epsilon_{\rm emit}$ is the emission coefficient and $\sigma_{\rm SB}=5.67\times 10^{-5}~\mbox{erg~cm$^{-2}$~K$^{-4}$~s$^{-1}$}$ is the Stefan--Boltzmann constant. 
The emission coefficient is \citep{Rizk+1991,Miura&Nakamoto2005}
\begin{eqnarray}
    \epsilon_{\rm emit} = \min{\left[ 2.1543\times 10^{-3} \left( \frac{a}{1~\mu\mbox{m}} \right)^{0.8253},1 \right]}.
    \label{eq:emission_coefficient}
\end{eqnarray}
The latent heat cooling by evaporation is the product of the evaporation rate, $J_{\rm evap}$, and the latent heat of evaporation, $L_{\rm evap}$,
\begin{eqnarray}
    \Lambda_{\rm evap} &=& J_{\rm evap} L_{\rm evap}, 
\end{eqnarray}
where $J_{\rm evap}$ is 
\begin{eqnarray}
    J_{\rm evap} &=& 691 \left( \frac{p_{\rm H_2}}{100~\mbox{dyn~cm}^{-2} } \right)^{1/2}
    \left( \frac{T }{ T_{\rm melt} } \right)^{-1/2} 
    \nonumber\\ &&\times 
    \exp{\left( -\frac{3.17\times10^4~\mbox{K}}{T} \right)} ~\mbox{g~cm$^{-2}$~s}^{-1},
    \label{eq:J_evap}
\end{eqnarray}
for forsterite \citep{Miura+2002,Miura&Nakamoto2005}, and $L_{\rm evap}=1.12\times 10^{11}~\mbox{erg~g}^{-1}$ \citep{Miura+2002}\footnote{
We note that the effect of the net gas-grain flow on the product of the evaporation rate, which is unclear, is not included. 
}.
The ambient gas pressure, $p_{\rm H_2}$, is the summation of the hydrostatic pressure and the ram pressure,
\begin{eqnarray}
    p_{\rm H_2} = n_{\rm g} k_{\rm B} T_{\rm g} + \frac{1}{3} m_{\rm g} n_{\rm g} (v -v_{\rm g})^2,
\end{eqnarray}
and the melting temperature is $T_{\rm melt}=2171~$K.
The latent heat cooling by evaporation becomes effective when the temperature of dust or a chondrule approaches the melting temperature.

Dust particles begin to melt when their temperature reaches the melting temperature.
According to \cite{Miura&Nakamoto2005}, we consider the phase transition of chondrules and dust between the solid and liquid. 
We express the phase transition by the solid (unmelted) mass fraction, $S$, 
\begin{eqnarray}
    m L_{\rm melt} \frac{{\rm d} S}{{\rm d} x} =
    \frac{4\pi a^2}{v} (\Gamma - \Lambda_{\rm rad} - \Lambda_{\rm evap} ),
\end{eqnarray}
where $L_{\rm melt}=4.47\times 10^9~\mbox{erg~g}^{-1}$ is the latent heat of melting.
During the phase transition ($0<S<1$), we assume that the temperature of a chondrule or dust remains constant and the net energy input is used for the phase transition. 
Molten dust particles crystallize according to the cooling. 
Droplets, which are dust particles with $S=0$, do not solidify immediately at $T_{\rm melt}$ since they cause supercooling \citep[e.g.,][]{Nagashima_Ken+2006, Nagashima_Ken+2008,Arakawa&Nakamoto2016a}.
In this study, we assume that the crystallization temperature of droplets is $T_{\rm c}=1000$~K.
For simplicity, we do not consider collisions between droplets that cause crystallization \citep[e.g.,][]{Arakawa&Nakamoto2016a}.

We consider the size evolution of dust particles. 
The size change of dust particles, $\left({{\rm d} a_{\rm d}}/{{\rm d} x}\right)$, is given by the summation of its contribution from boiling and evaporation:
\begin{equation}
\frac{{\rm d} a_{\rm d}}{{\rm d} x} = \left(\frac{{\rm d} a_{\rm d}}{{\rm d} x}\right)_{\rm boil} + \left(\frac{{\rm d} a_{\rm d}}{{\rm d} x}\right)_{\rm evap}.
\end{equation}
Dust particles boil when their vapor pressure, $p_{\rm eq}$, exceeds the gas pressure, $p_{\rm H_2}$ \citep{Miura+2002}, where 
\begin{eqnarray}
    p_{\rm eq} &=& 3.20 \times 10^8 \exp{\left(-\frac{6.18\times10^4~\mbox{K}}{T}\right)}~\mbox{bar}.
\end{eqnarray}
The boiling rate of a dust particle is 
\begin{eqnarray}
    -4\pi  a_{\rm d}^2 \rho_{\rm d} L_{\rm boil} \left(\frac{{\rm d} a_{\rm d}}{{\rm d} x}\right)_{\rm boil} =
    \frac{4\pi a_{\rm d}^2}{v_{\rm d}} (\Gamma - \Lambda_{\rm rad} - \Lambda_{\rm evap} ),
    \nonumber\\
\end{eqnarray}
where $L_{\rm boil}=1.64\times 10^{11}~\mbox{erg~g}^{-1}$ is the latent heat of boiling.
The size of a dust particle also decreases due to evaporation from the precursor surface, and its rate is 
\begin{eqnarray}
    -4\pi  a_{\rm d}^2 \rho_{\rm d} \left(\frac{{\rm d} a_{\rm d}}{{\rm d} x}\right)_{\rm evap} = \frac{4\pi a_{\rm d}^2}{v_{\rm d}} J_{\rm evap}.
\end{eqnarray}
\cite{Miura&Nakamoto2005} showed that evaporation is the main mechanism to shrink the dust size since the period of dust boiling is very short.
We do not consider the size evolution of chondrules since evaporation works more for smaller grains.

\subsection{Accretion of Rims}

Chondrules accrete droplets and/or crystallized them in the postshock region due to their relative velocities.
The accretion rate is 
\begin{eqnarray}
    \frac{{\rm d} m_{\rm ch} }{ {\rm d} x } &=& Q m_{\rm d} n_{\rm d} \pi a_{\rm ch}^2 \frac{v_{\rm imp}}{v_{\rm ch}},
\end{eqnarray}
where $Q$ is the accretion efficiency, $n_{\rm d}$ is the number density of the dust particles, and $v_{\rm imp}=|v_{\rm d}-v_{\rm ch}|$ is the impact velocity.
We simply consider that the number density of the dust particles is 
\begin{eqnarray}
    n_{\rm d} = \chi\frac{ m_{\rm g} n_{\rm g} }{ m_{\rm d} },
    \label{eq:n_d}
\end{eqnarray}
where the dust-to-gas ratio $\chi$ is set at 0.1. 
This is higher than 0.01 (the column dust-to-gas mass ratio in the minimum mass solar nebula \citep{Hayashi1981}), considering that chondrules and igneous rims are accreted in dusty environments \citep{Rubin2010, Schrader_DL+2013, Tenner+2015}.
We do not consider the evolution of $\chi$ behind the shock front. 
This is because the dust particles are expected to be well coupled to the gas at the location of accretion.
The final accreted rim thicknesses can be scaled by $\chi$.
We note that the realistic $\chi$ value is affected by dust velocity, accretion from chondrules, dust recondensation, dust evaporated from planetesimals \citep{Tanaka_KK+2013, Arakawa+2022} and disruption in collisions \citep[e.g.,][]{Jacquet&Thompson_C2014, Liffman2019}.
The growth rate of the rim thickness is given by
\begin{eqnarray}
    \frac{{\rm d} a_{\rm ch} }{ {\rm d} x } &=& \frac{1}{4\pi \rho_{\rm ch} a_{\rm ch}^2} \frac{{\rm d} m_{\rm ch} }{ {\rm d} x },
\end{eqnarray}
under the assumption that the density of the accreted rim is the same as $\rho_{\rm ch}$.
The average thickness of the rims, $\Delta$, is the difference between the chondrule size and the size of the chondrules with the accreted rims.

We consider the accretion criteria for molten dust particles and solid dust particles.
For molten dust particles, we consider the accretion criteria of the molten (or partially molten) particles on the solid chondrules.
The accretion condition between molten (or partially molten) particles is given by the Weber number, $\We$, which is the ratio of the impact energy to the surface energy, 
\begin{eqnarray}
    \We = \frac{2\rho_{\rm d} a v_{\rm imp}^2 }{\sigma},
\end{eqnarray}
where $\sigma=400~\mbox{erg~cm}^{-2}$ is the surface tension \citep{Murase&McBirney1973}.
According \cite{Mundo+1995}, we consider that molten dust particles are accreted when the Weber number is smaller than the critical Weber number for accretion, 
\begin{eqnarray}
    \We_{\rm cr} &=& \frac{9}{2} \beta^4 \Ca + 3 (1-\cos{\Theta})\beta^2 -12,
\end{eqnarray}
where $\beta\simeq2$--3 is the ratio of the maximum spreading diameter of the droplet to its initial diameter \citep{Chandra&Avedisian1991}, $\Theta$ is the contact angle, and $\Ca$ is the capillary number,
\begin{eqnarray}
    \Ca &=& \frac{\eta v_{\rm imp}}{ \sigma},
\end{eqnarray}
which is the function of the viscosity, $\eta$, \citep[][]{Hubbard2015}
\begin{eqnarray}
    \log_{10}{\left(\frac{\eta}{1~\mbox{P}}\right)} &=& -3.55 + \frac{5084.9~\mbox{K}}{T-584.9~\mbox{K}}.
    \label{eq:eta}
\end{eqnarray}
We ignore the dependence of the viscosity on the solid mass fraction \citep[e.g.,][]{Roscoe1952,Liu_Z+2018} since droplets are rapidly crystallized in the postshock region and the temperature dependence is stronger.
We simply adopt the maximum $\We_{\rm cr}$ values by taking $\beta=3$ and $\cos{\Theta}=-1$. 
These values do not affect results much since the critical Weber number is significantly affected by $\Ca$, i.e., $v_{\rm imp}$, and $T$.
We assume $Q=1$ when $\We<\We_{\rm cr}$, otherwise $Q=0$, for molten dust particles.

The accretion efficiency of solid dust particles is 
\begin{eqnarray}
    Q = 
    \left\{
        \begin{array}{l}
            Q_{\rm ad} \quad\hfill(v_{\rm min} \leq v_{\rm imp} \leq v_{\rm max} ),\\
           0 \hfill (\mbox{otherwise}),
        \end{array}
      \right.
      \label{eq:Q}
\end{eqnarray}
where $Q_{\rm ad}=0.5$, $v_{\rm min}=0.1~\mbox{km~s}^{-1}$, and $v_{\rm max}=1~\mbox{km~s}^{-1}$, taking from the fiducial case of \cite{Arakawa+2022}.
The velocity range of the accretion is based on the experimental studies, which consider the adhesion of ceramic particles on a ceramic substrate \citep[e.g.,][]{Hanft+2015}.
Dust particles with impact velocities less than $v_{\rm min}$ bounce and do not stick.
We ignore erosion in high-velocity collisions.
We do not consider collisions between dust particles.

\section{Results}\label{sect:results}

\subsection{Typical Accretion}\label{sect:typical}

\begin{figure*}[ht!]
    \plotone{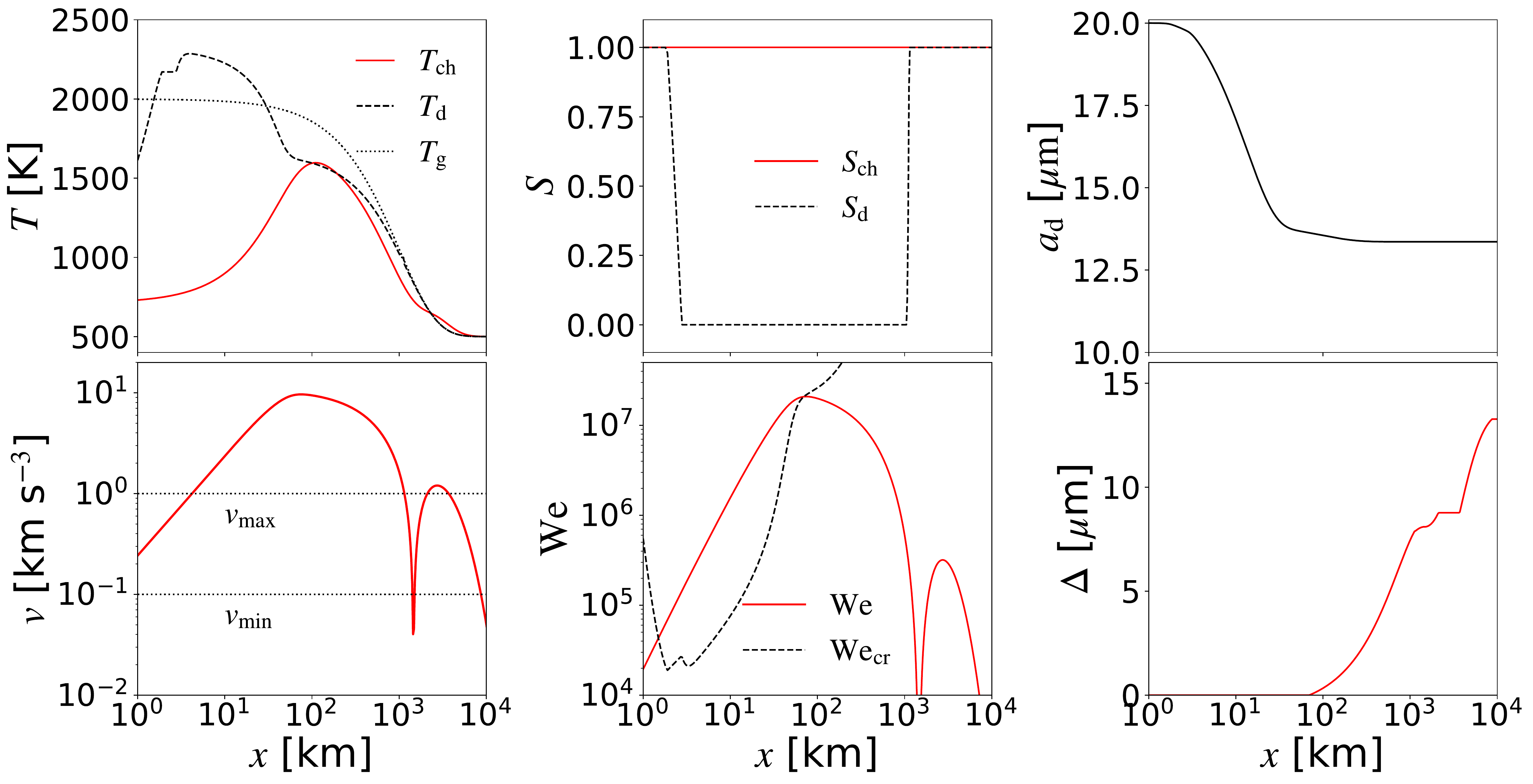}
    \caption{
        Temperature ($T$), solid (unmelted) mass fraction ($S$), dust size ($a_{\rm d}$), impact velocity ($v_{\rm imp}$), Weber number ($\We$), and rim thickness ($\Delta$) are plotted against distance from the shock front in our fiducial case, with $v_0=13~\mbox{km~s}^{-1}$, $n_0=10^{13.5}~\mbox{cm}^{-3}$, $L=10^3~\mbox{km}$, $a_{\rm ch}=10^3~\mbox{$\mu$m}$, and $a_{\rm d,init}=20~\mbox{$\mu$m}$.
        \label{fig:typical}}
\end{figure*}

We explain how igneous rims are accreted behind the shock front in our fiducial case, with $v_{\rm imp}=13~\mbox{km~s}^{-1}$, $n_0=10^{13.5}~\mbox{cm}^{-3}$, $L=10^3~\mbox{km}$, $a_{\rm ch}=10^3~\mbox{$\mu$m}$, and $a_{\rm d,init}=20~\mbox{$\mu$m}$. 
The temperature of the dust particles increases quickly and reaches the melting temperature (the top left panel of Figure \ref{fig:typical}).
The temperature of the dust particles is constant during melting, and it increases again after they become droplets ($S=0$, the top middle panel of Figure \ref{fig:typical}).
The temperature of the droplets becomes peaked at $x\simeq 4$~km when the heating rate is balanced with the cooling rate, which becomes higher as the temperature increases.
Droplets are cooled by evaporation, initially, and then radiation.
Droplets are crystallized at $T_{\rm c}$.
Through this evolution, the dust size shrinks and the final size is 13~$\mu$m (the top right panel of Figure \ref{fig:typical}).

Chondrules are also heated and they reach the peak temperature at $x\simeq 110$~km. 
The peak temperature is lower than the melting temperature.
This is because chondrules have a higher radiative cooling rate than dust since the emission coefficient increases as the size of the particle increases (Equation \ref{eq:emission_coefficient}).
Chondrules do not melt and $S_{\rm ch}$ is always 1.

The impact velocity between chondrules and dust particles initially increases (the bottom left panel of Figure \ref{fig:typical}).
Due to the stopping length dependence on the size, 
\begin{eqnarray}
    l_{\rm stop} &\simeq& 3.3\times 10^3~\mbox{km}
    \nonumber\\&&\times 
    \left( \frac{a}{1~\mbox{mm}} \right)
    \left( \frac{v}{v-v_{\rm g}} \right)^{2}
    \left( \frac{n_{\rm 0}}{10^{13.5}\mbox{~cm}^{-3}} \right)^{-1},
    \nonumber\\
    \label{eq:l_stop}
\end{eqnarray}
the relative velocity of a dust particle to the gas decreases, and that of a chondrule does not change.
The impact velocity becomes maximum, $v_{\rm imp}=9.6~\mbox{km~s}^{-1}$, which is close to $|v_{\rm post}-v_0|= 10.8~\mbox{km~s}^{-1}$, at $x\simeq 76$~km.
The impact velocity decreases as the relative velocity of a chondrule to gas decreases.
The impact velocity once reaches 0 but increases again \citep[][see also Appendix \ref{sect:v}]{Arakawa&Nakamoto2019, Arakawa+2022}.
The second peak impact velocity is $1.2~\mbox{km~s}^{-1}$ at $x\simeq 2.8\times 10^3~$km, and then the impact velocity decreases.

The Weber number and the critical Weber number are given by the temperature and impact velocity and are shown in the bottom middle panel of Figure \ref{fig:typical}.
The Weber number is larger than the critical Weber number during $T_{\rm d}>T_{\rm melt}$.
The Weber number peaks at $x\simeq 74$~km and begins to decrease, reflecting its $v_{\rm imp}$ dependence. 
The critical Weber number continues to increase due to its temperature dependence.
As a result, the Weber number becomes smaller than the critical Weber number at $x\simeq71$~km, where dust is in the supercooling state ($T_{\rm d}\simeq1.6\times 10^3$~K).
The condition of the droplet accretion ($\We < \We_{\rm cr}$) is satisfied in the supercooling state since 
\begin{eqnarray}
    T < &&
    584.9~\mbox{K} 
    \nonumber\\&&+
    \frac{ 5084.9~\mbox{K} }{ 5.07 + 
    \log{\left[
        \left( \frac{a_{\rm d}}{20~\mbox{$\mu$m}} \right)
        \left( \frac{v_{\rm imp} }{10~\mbox{km~s$^{-1}$}} \right)
    \right]} },
\end{eqnarray}
which is about 1600~K.
The droplet accretion occurs in $T_{\rm c}<T<T_{\rm cr}\simeq 1600~\mbox{K}$.
The droplet accretion stops when the supercooling droplets crystallize at $x=1.1\times10^3$~km.
Through the droplet accretion, chondrules acquire $7.9~\mu$m igneous rims (the bottom right panel of Figure \ref{fig:typical}).

The crystallized dust particles continue to be accreted onto chondrules.
When the dust particles crystallize at $x=1.1\times10^3$~km, the impact velocity is $1.0~\mbox{km~s}^{-1}$ and decreasing.
Dust particles are accreted except around the second peak impact velocity, where $v_{\rm imp}>v_{\rm max}$.
Finally, the thickness of an igneous rim becomes $13.3~\mu$m\footnote{
The rim thickness is less than the final dust size, indicating that the thickness is not uniform.
}.

We find that two stages of the accretion of igneous rims: droplet accretion and subsequent crystallized dust accretion (see Figure \ref{fig:schematic}).
The component dust particles of both layers experience a melting event.
The accretion of the crystallized dust particles occurs since chondrules still have the relative velocity to gas when the dust crystallize.
Considering $T_{\rm g}=T_{\rm d}$, which works well when the dust stopping length, $l_{\rm stop,d}$, is shorter than the shock scale, $L$, the dust crystallize when the dust temperature becomes $T_{\rm c}$ at $x = x_{\rm c}$, which is given by
\begin{eqnarray}
     T_{\rm g} &=& T_{\rm c} \Leftrightarrow x_{\rm c} = L \ln\left( \frac{T_{\rm post}-T_0}{T_{\rm c} - T_0} \right) \simeq L.
    \label{eq:x_sc}
\end{eqnarray}
When the relative velocity of chondrules is not totally damped at $L$, i.e., $l_{\rm stop,ch}\gtrsim L$, chondrules accrete crystallized dust particles.
The effect of the relation between $l_{\rm stop,ch}$ and $L$ is described in Section \ref{sect:L}.

\subsection{Parameter Dependence}

\subsubsection{Shock Velocity, Preshock Gas Number Density, and Initial Dust Size}

\begin{figure}[ht!]
    \plotone{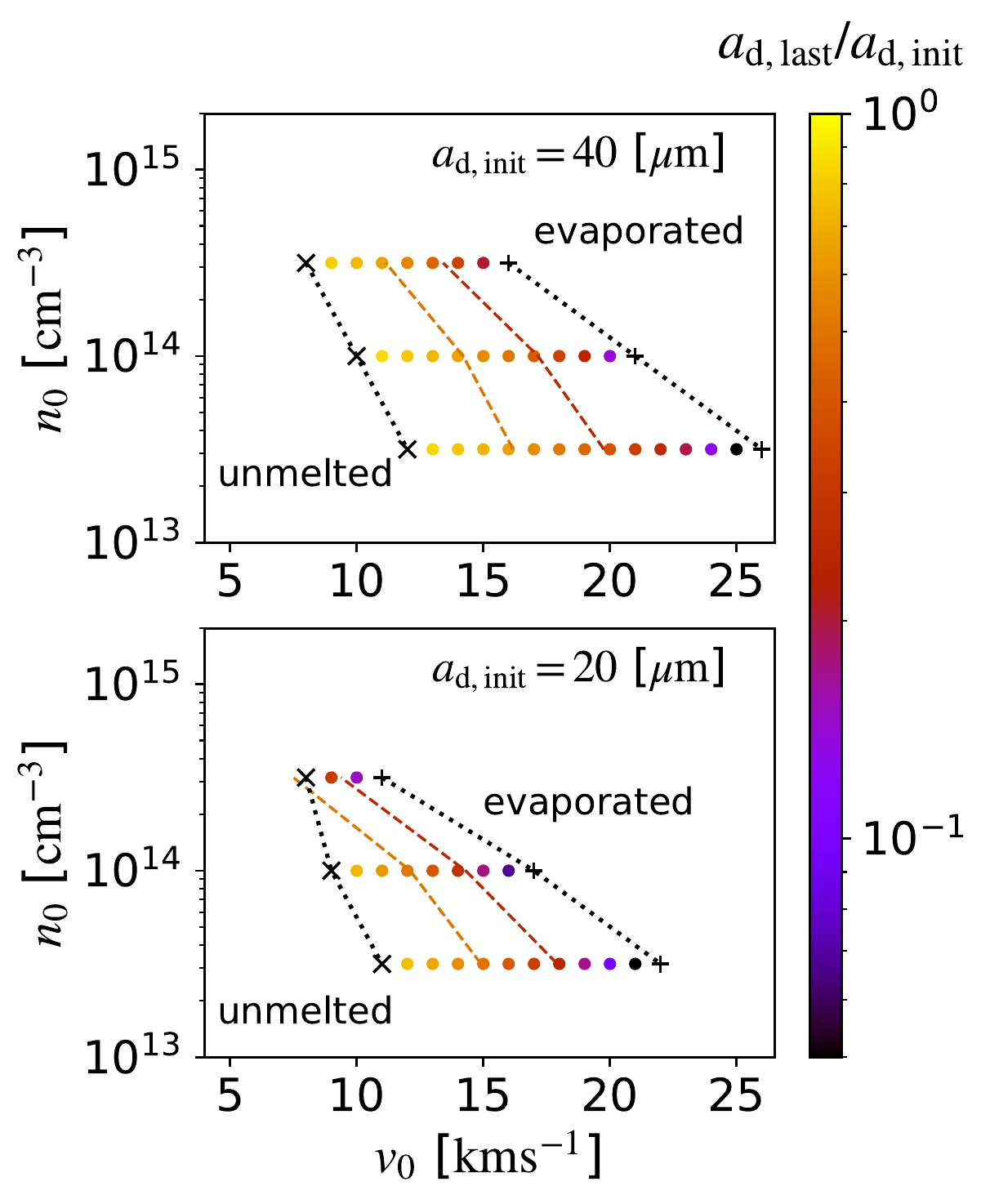}
    \caption{
        Condition for droplet formation on the shock velocity ($v_0$) and preshock gas number density ($n_0$) plane. 
        The top panel is the condition of $a_{\rm d,init}=40~\mu$m dust, and the bottom one is that of $a_{\rm d,init}=20~\mu$m dust.
        The other parameters are $L=10^3~\mbox{km}$ and $a_{\rm ch}=10^3~\mbox{$\mu$m}$.
        We plot cross symbols when the dust does not melt and plus symbols when the dust is completely evaporated.
        We plot circle symbols when dust particles become droplets and they do not evaporate completely.
        Their colors represent the size ratio between the final dust and initial dust ($a_{\rm d,last}/a_{\rm d,init}$).
        The dashed lines are plotted when the size ratios are 1/2 and 1/4.
        \label{fig:vs_n0_ad}}
\end{figure}

Igneous rims are accreted on chondrules in low-velocity shock waves.
In this section, we present when the igneous rim accretion occurs and how thick the rims are.
The accretion of igneous rims occurs when droplets are formed: the dust becomes droplets and does not evaporate completely.
These conditions depend on the shock velocity, preshock gas number density, and initial dust size \citep{Miura&Nakamoto2005}.
We plot the condition for the formation of droplets in Figure \ref{fig:vs_n0_ad}.
For example, the droplet formation occurs when $v_{0}$ is between $12~\mbox{km~s}^{-1}$ and $21~\mbox{km~s}^{-1}$ in the case that $n_0=10^{13.5}~\mbox{cm}^{-3}$, $L=10^3~\mbox{km}$, $a_{\rm ch}=10^3~\mbox{$\mu$m}$, and $a_{\rm d,init}=20~\mbox{$\mu$m}$.
Dust evaporates completely when the shock velocity is higher than this range, and dust does not melt when the shock velocity is lower.
Both boundary velocities of the melting and evaporation of dust become lower as the preshock gas number density increases.
This is because the heating rate via the energy transfer from gas is proportional to the preshock gas number density (Equation \ref{eq:Gamma}), although the radiative cooling rate is independent of the preshock gas number density (Equation \ref{eq:Lambda_rad}).
We note that chondrules also melt in high $n_0$ and/or $v_0$.

The shock velocity range for the droplet formation becomes wider as the initial dust size increases since a higher shock velocity is needed to evaporate completely.
In these shock velocity ranges, dust particles evaporate and their sizes shrink.
The accreted dust sizes are different according to the shock velocity, preshock gas number density, and initial dust size.
We plot the size ratios between the final and initial dust sizes, $a_{\rm d,last}/a_{\rm d,init}$ in Figure \ref{fig:vs_n0_ad}.
The dust sizes shrink more as the shock velocity and preshock gas number density are higher.
This suggests that if we consider a certain dust size, igneous rims are composed of larger grains as the disk gas dissipates when the shock velocity is the same.
The observed dust sizes of the components of the igneous rims are ranging from $\leq 3$--$100~\mu$m, and the averaged dust sizes are $\sim10~\mu$m in Allende and $\sim4~\mu$m in Semarkona \citep{Rubin1984}.
Such size distributions are reproduced in a single rim-forming shock event when the precursor dust particles have size distributions, where the precursor dust sizes are between the observed maximum size and the boundary size of the dust evaporation.

\begin{figure}[ht!]
    \plotone{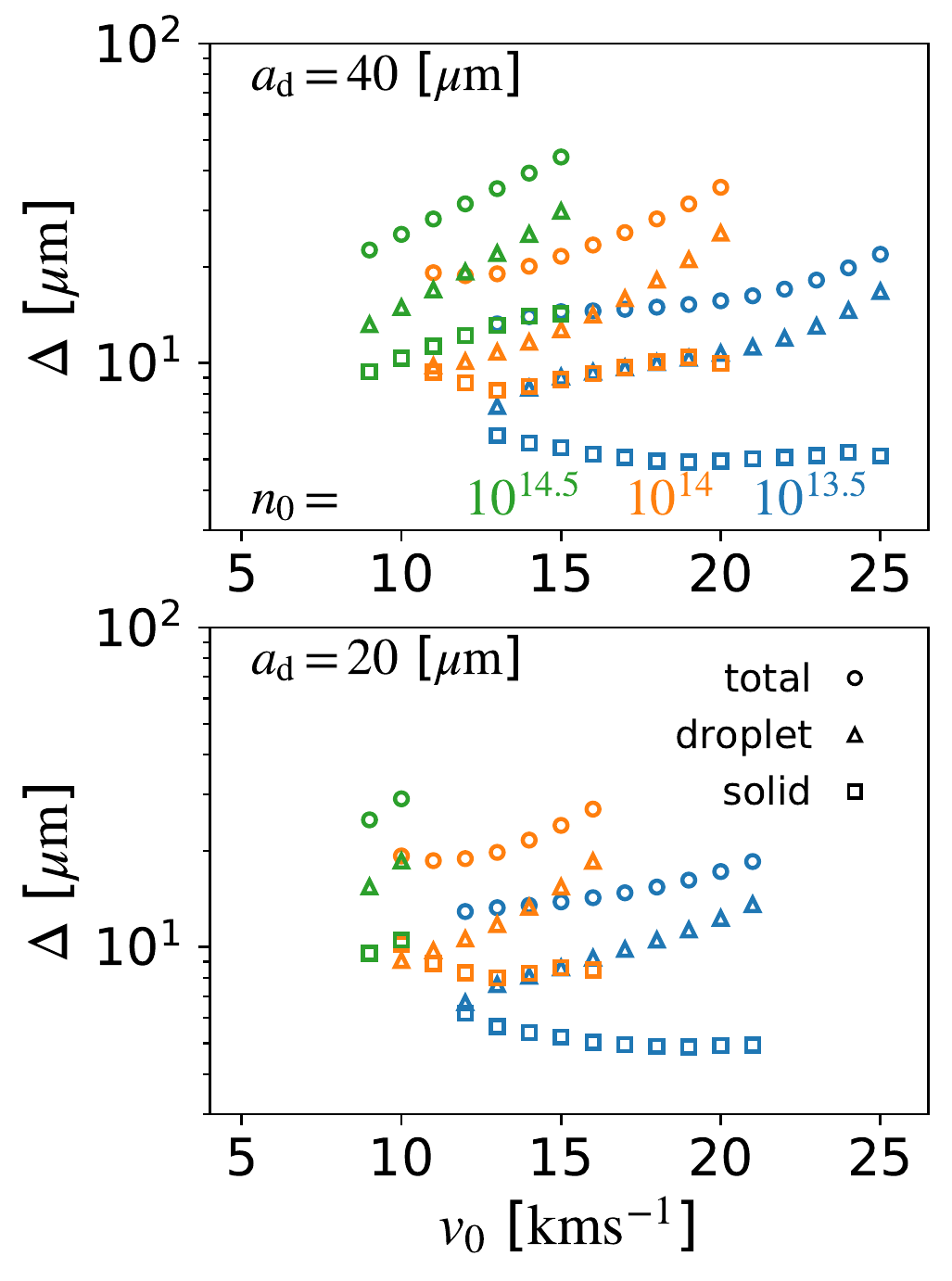}
    \caption{
        Thickness of the accreted rims ($\Delta$) is plotted as a function of the shock velocity. 
        The circle points are the total thicknesses of the rims, the triangles are the thicknesses of the accreted droplets, and the squares are the thicknesses of the accreted crystallized (solid) dust particles.
        The point colors represent the preshock gas number density, $n_0=10^{14.5}~\mbox{g~cm}^{-3}$ in green, $10^{14}~\mbox{g~cm}^{-3}$ in orange, and $10^{13.5}~\mbox{g~cm}^{-3}$ in blue.
        The top panel is the case of the $a_{\rm d,init}=40~\mu$m dust, and the bottom one is that of $a_{\rm d,init}=20~\mu$m dust.
        The other parameters are $L=10^3~\mbox{km}$ and $a_{\rm ch}=10^3~\mbox{$\mu$m}$.
        \label{fig:vs_arim_n0}}
\end{figure}

The thicknesses of the accreted rims are shown in Figure \ref{fig:vs_arim_n0}.
This figure shows that the thickness increases as the shock velocity increases.
This dependence is mainly because the dust number density increases as the shock velocity increases (Equations \ref{eq:n_g} and \ref{eq:n_d}).
This makes the thicknesses of the rims from the accretions of both droplets and crystallized particles larger.
However, the thicknesses of the rims from crystallized particles are not an increasing function of the shock velocity when the preshock gas number densities are $n_0 = 10^{13.5}~\mbox{g~cm}^{-3}$ and $10^{14}~\mbox{g~cm}^{-3}$.
In these cases, the impact velocity of the crystallized particles exceeds $v_{\rm max}$ {(Appendix \ref{sect:v}).
This makes the region, where the crystallized particles can be accreted by chondrules, shorter as the shock velocity increases.

The preshock gas number density also affects the rim thickness through the dust number density.
The rims from the accretion of the droplets become thicker as the preshock gas number density increases due to the dust number density.
In addition to the dust number density, the preshock gas number density affects the rim thickness of the crystallized particles through the length of the accretion region. 
We assume that the accretion region of the crystallized particles finishes when $v_{\rm imp}<v_{\rm min}$ (Equation \ref{eq:Q}).
When the gas number density is high, the stopping length becomes short (Equation \ref{eq:l_stop}), and the impact velocity becomes lower than $v_{\rm min}$, quickly (Appendix \ref{sect:v}).

The rim thicknesses are similar between the $a_{\rm d,init}=40~\mu$m cases and $a_{\rm d,init}=20~\mu$m cases.
This is because the precursor dust size mainly changes the peak locations of the dust temperature and impact velocity, which are much shorter than the accretion ranges of the droplets and crystallized dust.
This indicates that our results are good examples to consider the rim thicknesses that are composed of the observed coarse grains.

\subsubsection{Shock Scale}\label{sect:L}

\begin{figure}[ht!]
    \plotone{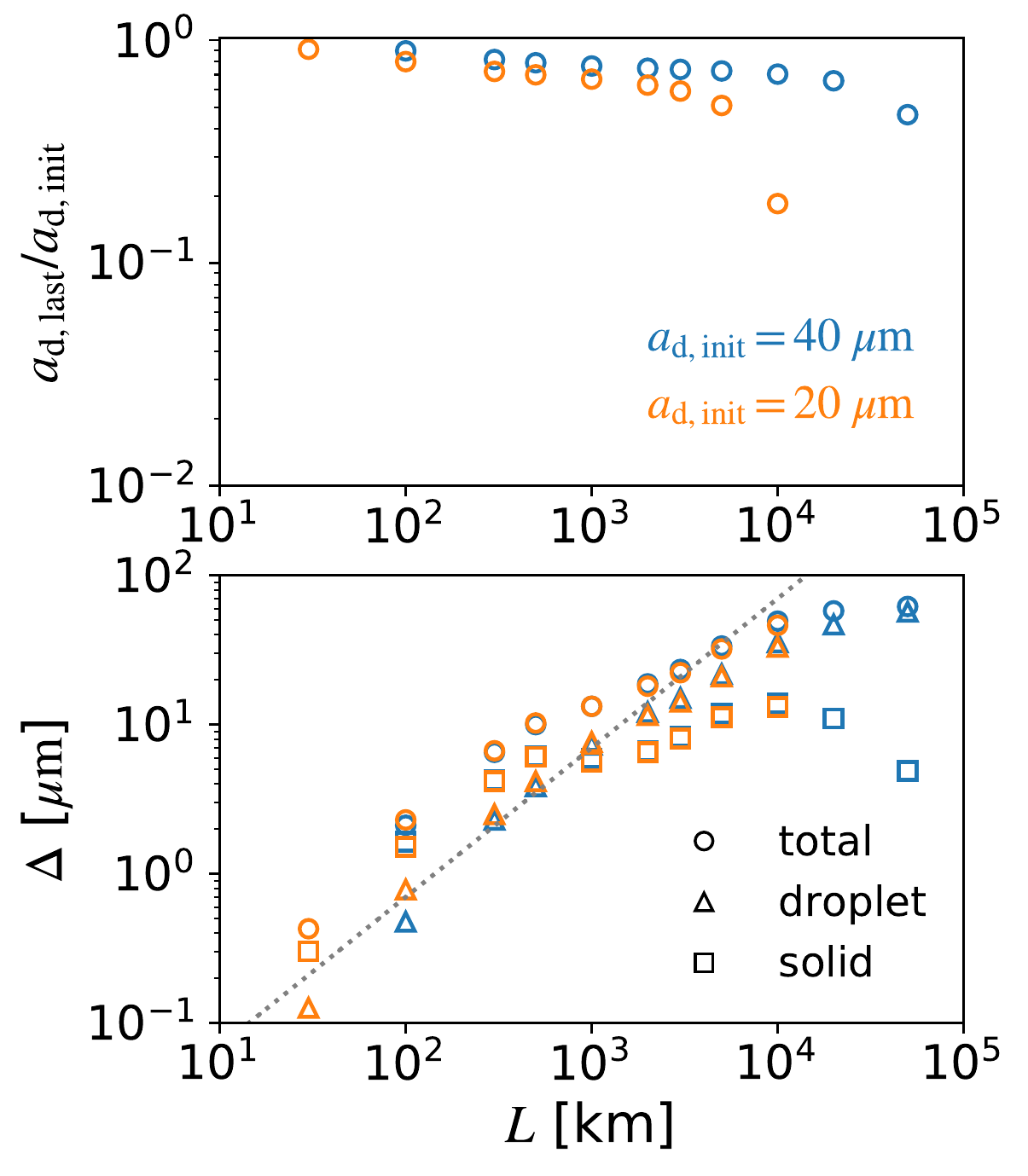}
    \caption{
        Size ratio between the final dust and initial dust ($a_{\rm d,last}/a_{\rm d,init}$, top) and thickness of the accreted rims ($\Delta$, bottom) are plotted as a function of the shock scale. 
        The point colors are different in the initial dust size, $a_{\rm d}=40~\mu$m in blue and $a_{\rm d}=20~\mu$m in orange.
        The points are plotted when the dust particles are molten but do not evaporate completely.
        The point styles in the bottom panel are the same as in Figure \ref{fig:vs_arim_n0}.
        The gray dotted line is a linear trend.
        The other parameters are the same as Figure \ref{fig:typical}.
        \label{fig:L_arim_ad}}
\end{figure}

The spatial scales of the shock waves are expected to depend on the shock-driving mechanisms, shock-driving bodies, and particle trajectories \citep[e.g.,][]{Morris_MA+2012, Desch+2012}.
We present how the shock scale affects the accretion condition and thickness of igneous rims.
Figure \ref{fig:L_arim_ad} shows the size ratio between the final dust and initial dust ($a_{\rm d,last}/a_{\rm d,init}$) and thickness of the accreted rims as a function of the shock scale in the $a_{\rm d,init}=40~\mu$m and $a_{\rm d,init}=20~\mu$m cases.
The other parameters are $v_0=13~\mbox{km~s}^{-1}$, $n_0=10^{13.5}~\mbox{cm}^{-3}$, and $a_{\rm ch}=10^3~\mbox{$\mu$m}$.
As the cases for the shock velocity and preshock gas number, dust does not melt completely when the shock scale is short, ($L<100$~km for $a_{\rm d,init}=40~\mu$m dust and $L<30$~km for $a_{\rm d,init}=20~\mu$m dust) and dust is evaporated when the shock scale is long 
 ($L>5\times 10^4$~km for $a_{\rm d,init}=40~\mu$m dust and $L>10^4$~km for $a_{\rm d,init}=20~\mu$m dust).
In the larger dust size case, a longer shock scale is needed for melting and evaporation.

The total thicknesses increase as the shock scale increases.
The droplets are accreted more as $L$ increases since the accretion of the droplets takes place at $x<x_{\rm c}\sim L$ (Equation \ref{eq:x_sc}).
The thickness of the accreted crystallized particles is also an increasing function of $L$ when $L< l_{\rm stop,ch}$.
This is because the accretion range of crystallized dust particles increases as $L$ increases.
The accretion of crystallized dust particles occurs between $x_{\rm c}\sim L$ and the location where the impact velocity becomes lower than $v_{\rm min}$ after the impact velocity reaches the second peak (Figure \ref{fig:schematic}, Section \ref{sect:typical}).
In the accretion of crystallized dust particles, the impact velocity is almost equal to the relative velocity of chondrules to gas in this range when $L\gtrsim 100~{\rm km}$ since crystallized dust particles are well coupled with gas.
When $L\ll l_{\rm stop,ch}$, the chondrule velocity barely changes behind the shock front since the gas velocity recovers to the preshock velocity at $l_{\rm stop,ch}$ \citep{Arakawa&Nakamoto2019}.
As $L$ increases, chondrules need longer $x$ to recover their velocities, which makes the accretion range of crystallized dust particles longer.
The thickness of the accreted crystallized particles is peaked around $L\sim l_{\rm stop,ch}$ and begins to decrease.
When $L\gg l_{\rm stop,ch}$, the chondrule velocity is close to the gas velocity at $x_{\rm c}$, and the accretion rate of crystallized dust particles decreases (Appendix \ref{sect:v}).

\subsubsection{Chondrule size}

\begin{figure}[ht!]
    \plotone{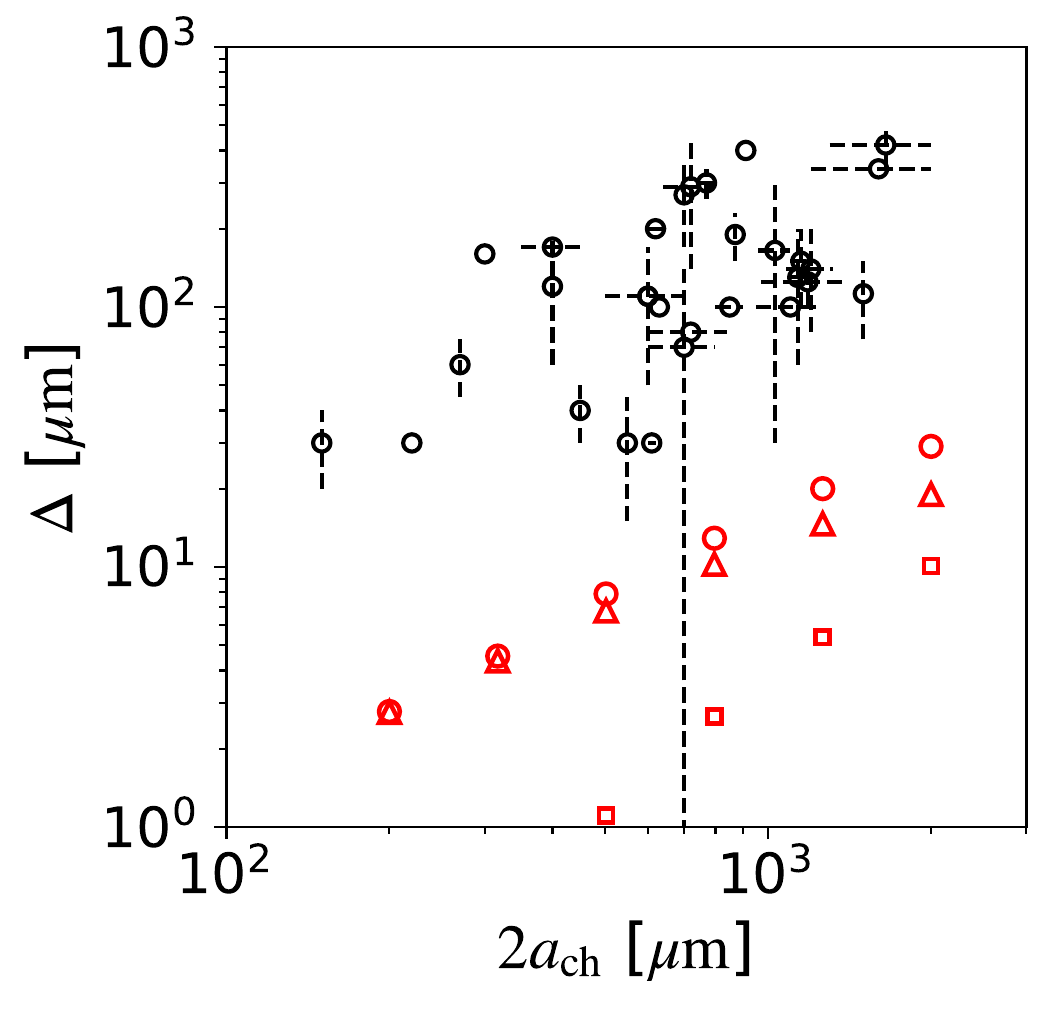}
    \caption{
        Thickness of the rims ($\Delta$) is plotted as a function of the chondrule diameter ($2a_{\rm ch}$).
        We plot our numerical results in the case that $v_0=10~\mbox{km~s}^{-1}$, $n_0=10^{14.5}~\mbox{cm}^{-3}$, $L=10^3~\mbox{km}$, and $a_{\rm d,init}=20~\mbox{$\mu$m}$ as red points.
        The red circles are the thicknesses of the total accreted dust, the red triangles are those of the droplets, and the red squares are those of the crystallized dust particles.
        The black circles are the thicknesses of the measured igneous rims and the dashed lines are the range of the minimum and maximum values \cite[][and the references therein, corrected mistakes in plotting some error bars]{Matsumoto+2021}.
        \label{fig:dch_arim_obs}}
\end{figure}

The observed igneous rims tend to increase as the chondrule size increases \citep{Matsumoto+2021}. 
This tendency is reproduced in our results.
Figure \ref{fig:dch_arim_obs} shows the thicknesses of the rims in the case that $v_0=10~\mbox{km~s}^{-1}$, $n_0=10^{14.5}~\mbox{cm}^{-3}$, $L=10^3~\mbox{km}$, and $a_{\rm d,init}=20~\mbox{$\mu$m}$.
As the chondrule size increases, the stopping length of chondrules becomes longer.
This makes high impact velocity accretion for droplets (Appendix \ref{sect:v}). 
Small-size chondrules do not have the rims of the accreted crystallized particles since $L>l_{\rm stop,ch}$ (Section \ref{sect:L}).

Although the observed igneous rims have a range in thickness, the observed rims are $\sim10$ times thicker than the rims in our results.
Multiple rim-forming shock events are needed to explain the thickness of the observed igneous rims, if the accretion behind the shock waves is the main source of the observed igneous rims.
We consider that multiple rim-forming shock events are plausible when chondrules form by shock events.
The rim-forming shock events are expected to be more frequent than the chondrule-forming shock events.
If a planetesimal generates a chondrule-forming bow shock, dust particles in higher impact parameters experience lower peak temperatures \citep{Boley+2013}.
Moreover, the shock velocity depends on its longitude \citep{Nagasawa+2019}.
These indicate that when a chondrule-forming bow shock is generated in a part of an orbit and at a certain impact parameter, the rim-forming bow shock is also generated in a wider part of the orbit and higher impact parameter, which means more chance to form igneous rims.
Such shock regions will help explain the depletion of volatile elements in the matrix \citep{Hubbard&Ebel2014}.
Igneous rims are accreted in multiple melting events as chondrules experience multiple (partial) melting events \citep[e.g.,][]{Rubin&Wasson1987,Baecker+2017}.

\section{Conclusion and Discussion}\label{sect:conclusion}

Igneous rims are chondrule-enveloping structures and are composed of coarse grains, which show evidence of melting.
We have explored whether igneous rims are accreted behind shock waves, which cause the melting of dust.
Our key findings are summarized as follows:
\begin{enumerate}
    \item We show that droplets, which are totally molten dust particles, are accreted onto chondrules.
    Droplets are accreted in their supercooling regime.
    Dust particles continue to be accreted after they are crystallized.
    These make two layers of rims. 
    \item Due to dust evaporation, the size of the accreted dust particles changes according to the shock velocity, preshock gas number density, and shock spatial scale. 
    As gas depletes, dust particles evaporate less and igneous rims become to be composed of larger grains at a certain velocity shock wave.
    \item The total thicknesses of the accreted rims increase as the shock velocity increases, the preshock gas number density increases, and the shock spatial scale increases.
    \item The thickness relation between the droplet layer and the crystallized-dust layer depends on the relation between the stopping length and the shock spatial scale, at which supercooling droplets are crystallized.
    \item To reproduce the thicknesses of the observed igneous rims, $\lesssim10$ rim-forming shock events are needed.
\end{enumerate}

We suggest that observed chondrules with igneous rims experience multiple shock events.
If chondrules with igneous rims experience multiple bow shock events, some of them, which have small impact parameters, reach melting temperature, are totally molten and form large single chondrules.
These large chondrules do not accrete droplets and crystallized dust since dust evaporates completely.
Some, which have large impact parameters, are not molten, accrete dust and become to have thicker igneous rims.
This scenario provides the reason why not all chondrules have igneous rims.

Behind the shock front, not only igneous texture dust but also fine-grained dust will be accreted onto chondrules.
There are some fine-grained dust forming mechanisms behind the shock front such as dust recondensation, evaporation of planetesimals \citep{Tanaka_KK+2013, Arakawa+2022}, and disruption from chondrules \citep[e.g.,][]{Jacquet&Thompson_C2014, Liffman2019}.
These are also accreted onto chondrules behind shock waves.
In addition, fine-grained dust is accreted onto chondrules in the solar nebula \citep[e.g.,][]{Metzler+1992, Matsumoto+2021}.
Such accreted dust will be transformed into igneous rims by shock waves or the other remelting events \citep[e.g.,][]{Rubin1984}.

In this study, we do not consider erosive collisions.
In the accretion of the crystallized dust particles, erosive collisions do not change our results since the impact velocity is lower than $v_{\rm max}$ in most cases.
The impacts between chondrules and droplets are initially $\We>\We_{\rm cr}$ and not accretional.
If these are erosive collisions, droplets would remove existing rims from the outer layers, which are composed of crystallized dust particles.
Such erosive droplet impacts make two-layer structures unclear.
Moreover, high-velocity collisions of droplets would cause chondrule erosion.
Droplet erosion plays a key role in the erosion of the rims and chondrules behind the shock waves.

\begin{acknowledgments}
    We are very grateful to the reviewers for their thoughtful comments
    We thank Prof. T. Nakamoto, Prof. H. Miura, and Dr. Kurosawa for suggestions and comments.
    Numerical simulations were in part carried out on analysis servers at Center for Computational Astrophysics, National Astronomical Observatory of Japan.
\end{acknowledgments}

\appendix
\section{Velocities of Chondrules and dust} \label{sect:v}

\begin{figure*}[ht!]
    \plotone{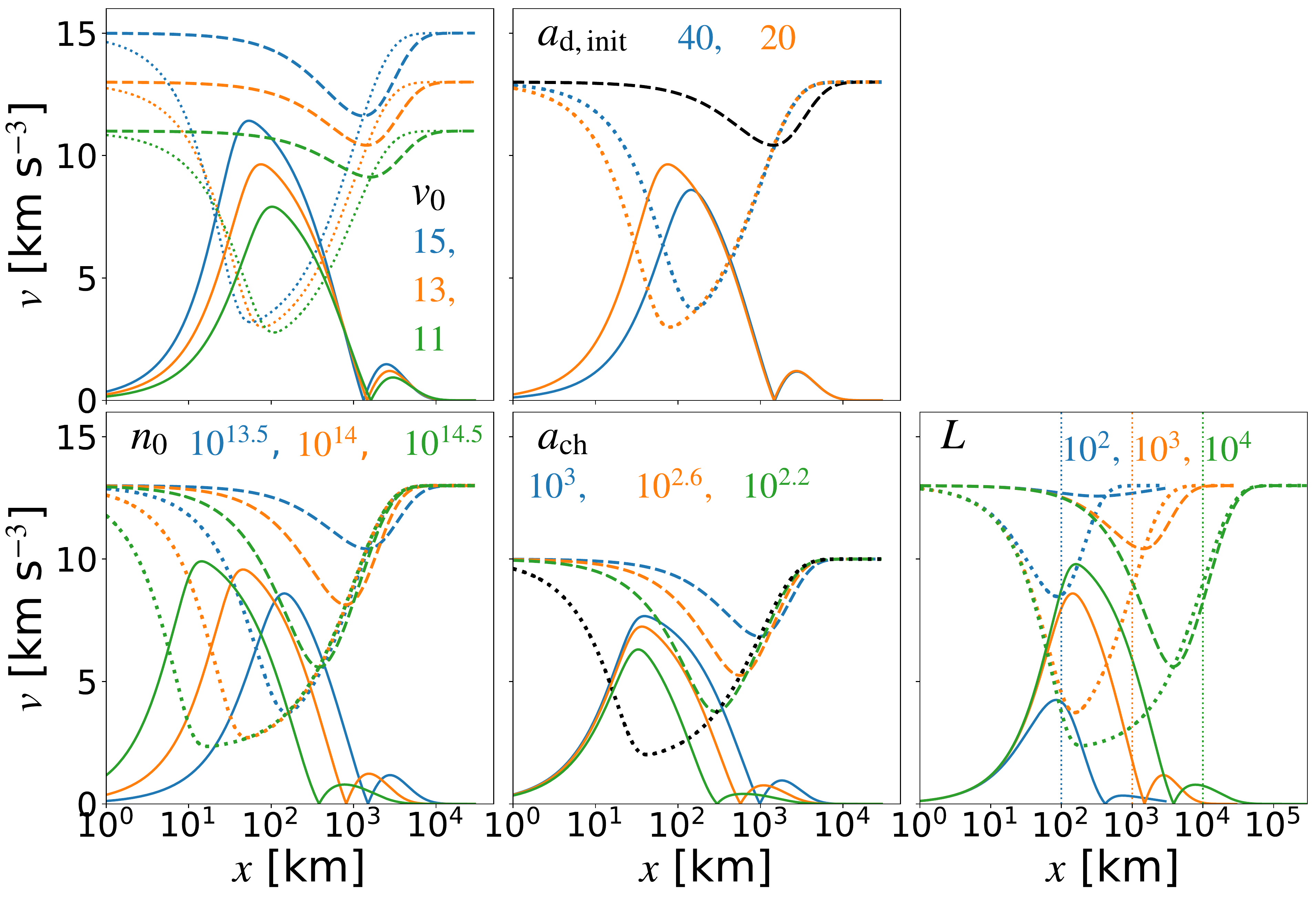}
    \caption{
    Evolution of the velocities of chondrules (dashed), dust (dotted), and impact (solid) is plotted as the function of $x$.
    In each panel, we show the parameter dependence on the velocity evolution: shock velocity ($v_0$, top-left), preshock gas number density ($n_0$, bottom-left), initial dust size ($a_{\rm d,init}$, top-middle), chondrules size ($a_{\rm ch}$, bottom-middle), and spatial scale of the shock ($L$, bottom-right). 
    The line colors are different between the parameters, and the lines are plotted in black when the lines are the same between the parameters.
    In the bottom-right panel, $L$ is plotted as the vertical dotted lines.
    The details are described in the text.
    \label{fig:x_v}}
\end{figure*}

The velocity evolution of chondrules and dust particles is important to consider for the accretion of crystallized dust particles.
Here, we show the evolution of velocities with changing parameters, shock velocity ($v_0$), preshock gas number density ($n_0$), initial dust size ($a_{\rm d,init}$),  chondrules size ($a_{\rm ch}$), and spatial scale of the shock ($L$).
The top-left panel of Figure \ref{fig:x_v} shows the dependence of the shock velocity ($v_0$).
The shock velocities are $15~{\rm km~s}^{-1}$ (blue), $13~{\rm km~s}^{-1}$ (orange), and $11~{\rm km~s}^{-1}$ (green), and the other parameters are $n_0=10^{13.5}~\mbox{cm}^{-3}$, $L=10^3~\mbox{km}$, $a_{\rm ch}=10^3~\mbox{$\mu$m}$, and $a_{\rm d,init}=20~\mbox{$\mu$m}$.
The shock velocity affects both velocities of chondrules and dust.
As the shock velocity increases, the peak impact velocity becomes higher since the chondrule velocity is $\simeq v_0$ and the dust velocity is $\simeq v_{\rm post}$ at the peak impact velocity (Section \ref{sect:typical}).
In contrast, the impact velocities are similar at $x>l_{\rm stop,ch}$.
The velocities of chondrules are relaxed to the gas velocity, and the impact velocity is hardly affected by the shock velocity in the accretion of crystallized dust particles.

The bottom-left panel shows the dependence of the preshock gas number density ($n_0$).
The preshock gas number densities are $10^{13.5}~\mbox{cm}^{-3}$ (blue), $10^{14}~\mbox{cm}^{-3}$ (orange), and $10^{14.5}~\mbox{cm}^{-3}$ (green), and the other parameters are $v_0=11~{\rm km~s}^{-1}$, $L=10^3~\mbox{km}$, $a_{\rm ch}=10^3~\mbox{$\mu$m}$, and $a_{\rm d,init}=40~\mbox{$\mu$m}$.
As the preshock gas number density decreases, the stopping lengths of chondrules and dust increase, which makes the impact velocity damped slowly.
The peak impact velocity increases as the preshock gas number density increases.
The second peak impact velocity becomes peaked at $L\sim l_{\rm stop,ch}$. 
When $n_0=10^{14}~\mbox{cm}^{-3}$ and $10^{14.5}~\mbox{cm}^{-3}$, the second peak impact velocities exceed $v_{\rm max}=1~{\rm km~s}^{-1}$.

The top-middle panel shows the dependence on the initial dust size ($a_{\rm d,init}$).
The initial dust sizes are $40~\mu$m (blue) and $20~\mu$m (orange), and the other parameters are $v_0=13~{\rm km~s}^{-1}$, $n_0=10^{13.5}~\mbox{cm}^{-3}$, $L=10^3~\mbox{km}$, and $a_{\rm ch}=10^3~\mbox{$\mu$m}$.
Smaller dust particles have shorter stopping lengths and their velocities quickly become the gas velocity.
After dust velocities are well coupled to the gas velocity, the impact velocities are the same.

The bottom-middle panel shows the dependence on the chondrules size ($a_{\rm ch}$).
The chondrule sizes are $10^3~\mu$m (blue), $10^{2.6}~\mu$m (orange), and $10^{2.2}~\mu$m (green), and the other parameters are $v_0=10~{\rm km~s}^{-1}$, $n_0=10^{14}~\mbox{cm}^{-3}$, $L=10^3~\mbox{km}$, and $a_{\rm d,init}=20~\mbox{$\mu$m}$.
Smaller chondrules quickly reach velocities comparable to the gas due to the short stopping length, resulting in a lower peak impact velocity.
In addition, smaller chondrules have lower second peaks of the impact velocities, which are quickly damped.

The bottom-right panel shows the dependence of the spatial scale of the shock ($L$). 
We note that the range of the horizontal axis in this panel is different from those in the other panels. 
The shock spatial scales are $10^2~$km (blue), $10^3~$km (orange), and $10^4~$km (green), and the other parameters are $v_0=13~{\rm km~s}^{-1}$, 
$n_0=10^{13.5}~\mbox{cm}^{-3}$, $a_{\rm ch}=10^3~\mbox{$\mu$m}$, and $a_{\rm d,init}=40~\mbox{$\mu$m}$.
The spatial shock scales change the gas velocity, which changes the velocities of chondrules and dust particles.
When $L=10^2$~km ($l_{\rm stop,d}<L<l_{\rm stop,ch}$), dust velocities are the same as the gas velocity at $x\ge L$. 
In this $L$ range, the gas velocity recovers to the preshock velocity before $x=l_{\rm stop,ch}$, and chondrule velocities barely change.
As $L$ increases, the gas velocity at $l_{\rm stop,d}$ decreases (Equation \ref{eq:v_g}). 
The impact velocity becomes higher since dust velocities are relaxed to the gas velocity at a lower velocity.
Moreover, chondrules velocities change more as $L$ increases, which makes the second peak structure of the impact velocity wider range.
When $L=10^3$~km ($L\simeq l_{\rm stop,ch}$), the impact velocity at the second peak becomes the highest. 
The second-peak velocity decreases as $L$ increases at $L>l_{\rm stop, ch}$ since the gas velocity does not change in the interval of $l_{\rm stop,ch}$.

\bibliography{bibtex_ym}{}

\begin{thebibliography}{}
\expandafter\ifx\csname natexlab\endcsname\relax\def\natexlab#1{#1}\fi
\providecommand{\url}[1]{\href{#1}{#1}}
\providecommand{\dodoi}[1]{doi:~\href{http://doi.org/#1}{\nolinkurl{#1}}}
\providecommand{\doeprint}[1]{\href{http://ascl.net/#1}{\nolinkurl{http://ascl.net/#1}}}
\providecommand{\doarXiv}[1]{\href{https://arxiv.org/abs/#1}{\nolinkurl{https://arxiv.org/abs/#1}}}

\bibitem[{{Arakawa} {et~al.}(2022){Arakawa}, {Kaneko}, \&
  {Nakamoto}}]{Arakawa+2022}
{Arakawa}, S., {Kaneko}, H., \& {Nakamoto}, T. 2022, \apj, 927, 188,
  \dodoi{10.3847/1538-4357/ac5254}

\bibitem[{{Arakawa} \& {Nakamoto}(2016)}]{Arakawa&Nakamoto2016a}
{Arakawa}, S., \& {Nakamoto}, T. 2016, \icarus, 276, 102,
  \dodoi{10.1016/j.icarus.2016.04.041}

\bibitem[{{Arakawa} \& {Nakamoto}(2019)}]{Arakawa&Nakamoto2019}
---. 2019, \apj, 877, 84, \dodoi{10.3847/1538-4357/ab1b3e}

\bibitem[{{Ashworth}(1977)}]{Ashworth1977}
{Ashworth}, J.~R. 1977, Earth and Planetary Science Letters, 35, 25,
  \dodoi{10.1016/0012-821X(77)90024-3}

\bibitem[{{Asphaug} {et~al.}(2011){Asphaug}, {Jutzi}, \&
  {Movshovitz}}]{Asphaug+2011}
{Asphaug}, E., {Jutzi}, M., \& {Movshovitz}, N. 2011, Earth and Planetary
  Science Letters, 308, 369, \dodoi{10.1016/j.epsl.2011.06.007}

\bibitem[{{Baecker} {et~al.}(2017){Baecker}, {Rubin}, \&
  {Wasson}}]{Baecker+2017}
{Baecker}, B., {Rubin}, A.~E., \& {Wasson}, J.~T. 2017, \gca, 211, 256,
  \dodoi{10.1016/j.gca.2017.05.013}

\bibitem[{{Boley} {et~al.}(2013){Boley}, {Morris}, \& {Desch}}]{Boley+2013}
{Boley}, A.~C., {Morris}, M.~A., \& {Desch}, S.~J. 2013, \apj, 776, 101,
  \dodoi{10.1088/0004-637X/776/2/101}

\bibitem[{{Chandra} \& {Avedisian}(1991)}]{Chandra&Avedisian1991}
{Chandra}, S., \& {Avedisian}, C.~T. 1991, Proceedings of the Royal Society of
  London Series A, 432, 13, \dodoi{10.1098/rspa.1991.0002}

\bibitem[{{Ciesla}(2006)}]{Ciesla2006}
{Ciesla}, F.~J. 2006, \maps, 41, 1347,
  \dodoi{10.1111/j.1945-5100.2006.tb00526.x}

\bibitem[{{Desch} {et~al.}(2012){Desch}, {Morris}, {Connolly}, \&
  {Boss}}]{Desch+2012}
{Desch}, S.~J., {Morris}, M.~A., {Connolly}, H.~C., \& {Boss}, A.~P. 2012,
  \maps, 47, 1139, \dodoi{10.1111/j.1945-5100.2012.01357.x}

\bibitem[{{Friedrich} {et~al.}(2015){Friedrich}, {Weisberg}, {Ebel}, {Biltz},
  {Corbett}, {Iotzov}, {Khan}, \& {Wolman}}]{Friedrich+2015}
{Friedrich}, J.~M., {Weisberg}, M.~K., {Ebel}, D.~S., {et~al.} 2015, Chemie der
  Erde / Geochemistry, 75, 419, \dodoi{10.1016/j.chemer.2014.08.003}

\bibitem[{{Gombosi} {et~al.}(1986){Gombosi}, {Nagy}, \&
  {Cravens}}]{Gombosi+1986}
{Gombosi}, T.~I., {Nagy}, A.~F., \& {Cravens}, T.~E. 1986, Reviews of
  Geophysics, 24, 667, \dodoi{10.1029/RG024i003p00667}

\bibitem[{{Hanft} {et~al.}(2015){Hanft}, {Exner}, {Schubert}, {St{\"o}cker},
  {Fuierer}, \& {Moos}}]{Hanft+2015}
{Hanft}, D., {Exner}, J., {Schubert}, M., {et~al.} 2015, Journal of Ceramic
  Science and Technology, 6, 147, \dodoi{10.4416/JCST2015-00018}

\bibitem[{{Hanna} \& {Ketcham}(2018)}]{Hanna&Ketcham2018}
{Hanna}, R.~D., \& {Ketcham}, R.~A. 2018, Earth and Planetary Science Letters,
  481, 201, \dodoi{10.1016/j.epsl.2017.10.029}

\bibitem[{{Hayashi}(1981)}]{Hayashi1981}
{Hayashi}, C. 1981, Progress of Theoretical Physics Supplement, 70, 35,
  \dodoi{10.1143/PTPS.70.35}

\bibitem[{{Hewins} {et~al.}(2005){Hewins}, {Connolly}, {Lofgren}, \&
  {Libourel}}]{Hewins+2005}
{Hewins}, R.~H., {Connolly}, H.~C., {Lofgren}, G.~E., J., \& {Libourel}, G.
  2005, in Astronomical Society of the Pacific Conference Series, Vol. 341,
  Chondrites and the Protoplanetary Disk, ed. A.~N. {Krot}, E.~R.~D. {Scott},
  \& B.~{Reipurth}, 286

\bibitem[{{Hood} \& {Horanyi}(1991)}]{Hood&Horanyi1991}
{Hood}, L.~L., \& {Horanyi}, M. 1991, \icarus, 93, 259,
  \dodoi{10.1016/0019-1035(91)90211-B}

\bibitem[{{Hor{\'a}nyi} {et~al.}(1995){Hor{\'a}nyi}, {Morfill}, {Goertz}, \&
  {Levy}}]{Horanyi+1995}
{Hor{\'a}nyi}, M., {Morfill}, G., {Goertz}, C.~K., \& {Levy}, E.~H. 1995,
  \icarus, 114, 174, \dodoi{10.1006/icar.1995.1052}

\bibitem[{{Hubbard}(2015)}]{Hubbard2015}
{Hubbard}, A. 2015, \icarus, 254, 56, \dodoi{10.1016/j.icarus.2015.02.030}

\bibitem[{{Hubbard} \& {Ebel}(2014)}]{Hubbard&Ebel2014}
{Hubbard}, A., \& {Ebel}, D.~S. 2014, \icarus, 237, 84,
  \dodoi{10.1016/j.icarus.2014.04.015}

\bibitem[{{Hughes}(1978)}]{Hughes_DW1978}
{Hughes}, D.~W. 1978, Earth and Planetary Science Letters, 38, 391,
  \dodoi{10.1016/0012-821X(78)90113-9}

\bibitem[{{Iida} {et~al.}(2001){Iida}, {Nakamoto}, {Susa}, \&
  {Nakagawa}}]{Iida+2001}
{Iida}, A., {Nakamoto}, T., {Susa}, H., \& {Nakagawa}, Y. 2001, \icarus, 153,
  430, \dodoi{10.1006/icar.2001.6682}

\bibitem[{{Jacquet} {et~al.}(2015){Jacquet}, {Alard}, \&
  {Gounelle}}]{Jacquet+2015}
{Jacquet}, E., {Alard}, O., \& {Gounelle}, M. 2015, \gca, 155, 47,
  \dodoi{10.1016/j.gca.2015.02.005}

\bibitem[{{Jacquet} {et~al.}(2013){Jacquet}, {Paulhiac-Pison}, {Alard},
  {Kearsley}, \& {Gounelle}}]{Jacquet+2013}
{Jacquet}, E., {Paulhiac-Pison}, M., {Alard}, O., {Kearsley}, A.~T., \&
  {Gounelle}, M. 2013, \maps, 48, 1981, \dodoi{10.1111/maps.12212}

\bibitem[{{Jacquet} \& {Thompson}(2014)}]{Jacquet&Thompson_C2014}
{Jacquet}, E., \& {Thompson}, C. 2014, \apj, 797, 30,
  \dodoi{10.1088/0004-637X/797/1/30}

\bibitem[{{Johansen} \& {Okuzumi}(2018)}]{Johansen&Okuzumi2018}
{Johansen}, A., \& {Okuzumi}, S. 2018, \aap, 609, A31,
  \dodoi{10.1051/0004-6361/201630047}

\bibitem[{{Johnson} {et~al.}(2015){Johnson}, {Minton}, {Melosh}, \&
  {Zuber}}]{Johnson+2015}
{Johnson}, B.~C., {Minton}, D.~A., {Melosh}, H.~J., \& {Zuber}, M.~T. 2015,
  \nat, 517, 339, \dodoi{10.1038/nature14105}

\bibitem[{{Liffman}(2019)}]{Liffman2019}
{Liffman}, K. 2019, Geochimica et Cosmochimica Acta, 264, 118,
  \dodoi{10.1016/j.gca.2019.08.009}

\bibitem[{{Liu} {et~al.}(2018){Liu}, {Pandelaers}, {Blanpain}, \&
  {Guo}}]{Liu_Z+2018}
{Liu}, Z., {Pandelaers}, L., {Blanpain}, B., \& {Guo}, M. 2018, Metallurgical
  and Materials Transactions B, 49, 2469, \dodoi{10.1007/s11663-018-1374-9}

\bibitem[{{Matsuda} {et~al.}(2019){Matsuda}, {Sakamoto}, {Tachibana}, \&
  {Yurimoto}}]{Matsuda+2019}
{Matsuda}, N., {Sakamoto}, N., {Tachibana}, S., \& {Yurimoto}, H. 2019, Chemie
  der Erde / Geochemistry, 79, 125524, \dodoi{10.1016/j.chemer.2019.07.006}

\bibitem[{{Matsumoto} {et~al.}(2021){Matsumoto}, {Hasegawa}, {Matsuda}, \&
  {Liu}}]{Matsumoto+2021}
{Matsumoto}, Y., {Hasegawa}, Y., {Matsuda}, N., \& {Liu}, M.-C. 2021, \icarus,
  367, 114538, \dodoi{10.1016/j.icarus.2021.114538}

\bibitem[{{Metzler} {et~al.}(1992){Metzler}, {Bischoff}, \&
  {Stoeffler}}]{Metzler+1992}
{Metzler}, K., {Bischoff}, A., \& {Stoeffler}, D. 1992, Geochimica et
  Cosmochimica Acta, 56, 2873, \dodoi{10.1016/0016-7037(92)90365-P}

\bibitem[{{Miura} \& {Nakamoto}(2005)}]{Miura&Nakamoto2005}
{Miura}, H., \& {Nakamoto}, T. 2005, \icarus, 175, 289,
  \dodoi{10.1016/j.icarus.2004.11.011}

\bibitem[{{Miura} {et~al.}(2002){Miura}, {Nakamoto}, \& {Susa}}]{Miura+2002}
{Miura}, H., {Nakamoto}, T., \& {Susa}, H. 2002, \icarus, 160, 258,
  \dodoi{10.1006/icar.2002.6964}

\bibitem[{{Morris} {et~al.}(2012){Morris}, {Boley}, {Desch}, \&
  {Athanassiadou}}]{Morris_MA+2012}
{Morris}, M.~A., {Boley}, A.~C., {Desch}, S.~J., \& {Athanassiadou}, T. 2012,
  \apj, 752, 27, \dodoi{10.1088/0004-637X/752/1/27}

\bibitem[{{Mundo} {et~al.}(1995){Mundo}, {Sommerfeld}, \&
  {Tropea}}]{Mundo+1995}
{Mundo}, C.~H.~R., {Sommerfeld}, M., \& {Tropea}, C. 1995, International
  journal of multiphase flow, 21, 151, \dodoi{10.1016/0301-9322(94)00069-V}

\bibitem[{{Murase} \& {McBirney}(1973)}]{Murase&McBirney1973}
{Murase}, T., \& {McBirney}, A.~R. 1973, Geological Society of America
  Bulletin, 84, 3563, \dodoi{10.1130/0016-7606(1973)84<3563:POSCIR>2.0.CO;2}

\bibitem[{{Nagasawa} {et~al.}(2019){Nagasawa}, {Tanaka}, {Tanaka}, {Nomura},
  {Nakamoto}, \& {Miura}}]{Nagasawa+2019}
{Nagasawa}, M., {Tanaka}, K.~K., {Tanaka}, H., {et~al.} 2019, \apj, 871, 110,
  \dodoi{10.3847/1538-4357/aaf795}

\bibitem[{{Nagashima} {et~al.}(2008){Nagashima}, {Moriuchi}, {Tsukamoto},
  {Tanaka}, \& {Kobatake}}]{Nagashima_Ken+2008}
{Nagashima}, K., {Moriuchi}, Y., {Tsukamoto}, K., {Tanaka}, K.~K., \&
  {Kobatake}, H. 2008, Journal of Mineralogical and Petrological Sciences, 103,
  204, \dodoi{10.2465/jmps.070620c}

\bibitem[{{Nagashima} {et~al.}(2006){Nagashima}, {Tsukamoto}, {Satoh},
  {Kobatake}, \& {Dold}}]{Nagashima_Ken+2006}
{Nagashima}, K., {Tsukamoto}, K., {Satoh}, H., {Kobatake}, H., \& {Dold}, P.
  2006, Journal of Crystal Growth, 293, 193,
  \dodoi{10.1016/j.jcrysgro.2006.01.064}

\bibitem[{{Nakamoto} \& {Miura}(2004)}]{Nakamoto&Miura2004_LPSC}
{Nakamoto}, T., \& {Miura}, H. 2004, in Lunar and Planetary Science Conference,
  ed. S.~{Mackwell} \& E.~{Stansbery}, Lunar and Planetary Science Conference,
  1847

\bibitem[{{Pinto} {et~al.}(2022){Pinto}, {Marrocchi}, {Jacquet}, \&
  {Olivares}}]{Pinto+2022}
{Pinto}, G.~A., {Marrocchi}, Y., {Jacquet}, E., \& {Olivares}, F. 2022, \maps,
  57, 1004, \dodoi{10.1111/maps.13812}

\bibitem[{{Rizk} {et~al.}(1991){Rizk}, {Hunten}, \& {Engel}}]{Rizk+1991}
{Rizk}, B., {Hunten}, D.~M., \& {Engel}, S. 1991, \jgr, 96, 1303,
  \dodoi{10.1029/90JA01998}

\bibitem[{{Roscoe}(1952)}]{Roscoe1952}
{Roscoe}, R. 1952, British Journal of Applied Physics, 3, 267,
  \dodoi{10.1088/0508-3443/3/8/306}

\bibitem[{{Rubin}(1984)}]{Rubin1984}
{Rubin}, A.~E. 1984, Geochimica et Cosmochimica Acta, 48, 1779,
  \dodoi{10.1016/0016-7037(84)90032-2}

\bibitem[{{Rubin}(2010)}]{Rubin2010}
---. 2010, Geochimica et Cosmochimica Acta, 74, 4807,
  \dodoi{10.1016/j.gca.2010.05.018}

\bibitem[{{Rubin} \& {Wasson}(1987)}]{Rubin&Wasson1987}
{Rubin}, A.~E., \& {Wasson}, J.~T. 1987, \gca, 51, 1923,
  \dodoi{10.1016/0016-7037(87)90182-7}

\bibitem[{{Schrader} {et~al.}(2013){Schrader}, {Connolly}, {Lauretta},
  {Nagashima}, {Huss}, {Davidson}, \& {Domanik}}]{Schrader_DL+2013}
{Schrader}, D.~L., {Connolly}, H.~C., {Lauretta}, D.~S., {et~al.} 2013, \gca,
  101, 302, \dodoi{10.1016/j.gca.2012.09.045}

\bibitem[{{Scott}(2007)}]{Scott2007}
{Scott}, E. R.~D. 2007, Annual Review of Earth and Planetary Sciences, 35, 577,
  \dodoi{10.1146/annurev.earth.35.031306.140100}

\bibitem[{{Susa} {et~al.}(1998){Susa}, {Uehara}, {Nishi}, \&
  {Yamada}}]{Susa+1998}
{Susa}, H., {Uehara}, H., {Nishi}, R., \& {Yamada}, M. 1998, Progress of
  Theoretical Physics, 100, 63, \dodoi{10.1143/PTP.100.63}

\bibitem[{{Tanaka} {et~al.}(2013){Tanaka}, {Yamamoto}, {Tanaka}, {Miura},
  {Nagasawa}, \& {Nakamoto}}]{Tanaka_KK+2013}
{Tanaka}, K.~K., {Yamamoto}, T., {Tanaka}, H., {et~al.} 2013, \apj, 764, 120,
  \dodoi{10.1088/0004-637X/764/2/120}

\bibitem[{{Tenner} {et~al.}(2015){Tenner}, {Nakashima}, {Ushikubo}, {Kita}, \&
  {Weisberg}}]{Tenner+2015}
{Tenner}, T.~J., {Nakashima}, D., {Ushikubo}, T., {Kita}, N.~T., \& {Weisberg},
  M.~K. 2015, \gca, 148, 228, \dodoi{10.1016/j.gca.2014.09.025}

\bibitem[{{Wakita} {et~al.}(2017){Wakita}, {Matsumoto}, {Oshino}, \&
  {Hasegawa}}]{Wakita+2017}
{Wakita}, S., {Matsumoto}, Y., {Oshino}, S., \& {Hasegawa}, Y. 2017, \apj, 834,
  125, \dodoi{10.3847/1538-4357/834/2/125}

\bibitem[{{Wasson}(1993)}]{Wasson1993}
{Wasson}, J.~T. 1993, Meteoritics, 28, 14,
  \dodoi{10.1111/j.1945-5100.1993.tb00244.x}

\end{thebibliography}
\bibliographystyle{aasjournal}

\end{document}